\documentclass[11pt,english]{article}
\usepackage[T1]{fontenc}
\usepackage[english]{babel}

\usepackage[top=1in,bottom=1in,left=0.75in,right=0.75in]{geometry}
\usepackage{times}
\usepackage{amsmath,amssymb,amsfonts}
\usepackage{graphicx}

\usepackage{natbib}
\usepackage{booktabs}
\usepackage{caption}
\usepackage{subcaption}
\usepackage{float}

\usepackage{hyperref}
\hypersetup{
  colorlinks   = true,
  linkcolor    = blue,
  citecolor    = blue,
  urlcolor     = cyan
}

\usepackage{titlesec}
\titleformat{\section}
  {\normalfont\large\bfseries}{\thesection.}{1em}{}
\titleformat{\subsection}
  {\normalfont\normalsize\bfseries}{\thesubsection}{1em}{}
\titlespacing*{\section}{0pt}{12pt}{6pt}
\titlespacing*{\subsection}{0pt}{12pt}{6pt}

\usepackage[flushmargin]{footmisc}

\usepackage[stretch=10,shrink=10,step=1]{microtype}
\usepackage{enumitem}

\usepackage{tabularx}
\newcolumntype{L}{>{\raggedright\arraybackslash}X}
\begin{document}
\twocolumn[
  \begin{center}
    {\LARGE Introducing Powerwise (PWR): A pairwise and Power Rating method for selecting at‑large teams to the NCAA Division I Men's Lacrosse Championship \par}
    \vskip 1.5em
    {\large
      \lineskip 2em
      \begin{tabular}[t]{c}
        Lawrence Feldman$^{1}$ \\
        \texttt{laf@laxnumbers.com}
        \and
        Matthew Bomparola$^{2}$ \\
        \texttt{matthewbomparola@gmail.com}
      \end{tabular}\par}
    \vskip 1.5em
    Contributors:\\\textit{The Men's Division I Lacrosse Selection Criteria and Ranking Committee (SCR),\\ chaired by UVA Head Coach Lars Tiffany$^{3}$.}
  \end{center}
  
  \vskip 1.5em
  \noindent\textbf{Abstract:} This document describes a new system for selecting teams for the NCAA Men’s Division I Lacrosse championship tournament called “Powerwise” that was developed in discussions with the NCAA Lacrosse Selection Criteria and Ranking Committee (SCR). The method is simple, employing hierarchical pairwise comparisons that emphasize on-field performance in head-to-head and common opponent matchups and a simple Massey/Colley/Sagarin-like statistic called the Power Rating (PR) when on-field results are not conclusive. Power Ratings are based on margin of victory and implicitly account for strength of schedule. Powerwise addresses the complexities of team selection in a way that both coaches and fans can understand while improving the fairness, objectivity, and overall quality of the selection process.
  
  \vskip 0.5em
  \noindent\textbf{Keywords:} At-large selection; men’s Division I lacrosse; pairwise comparisons; Power Ratings; goal differences; on-field results; Sagarin; Massey; Colley. 
  
  \vskip 2em
]

\footnotetext[1]{Lawrence Feldman played lacrosse at the University of Pennsylvania, received a PhD from Auburn University and attended Stanford University as a Post Doc. He worked as an engineer/computer analyst in the USAF and worked in scientific laboratories, including Lawrence Livermore Laboratory (LLL) and Los Alamos National Laboratory (LANL). In addition, he worked for NASA and was a software architect for Intel Corp. Feldman has served on the Coaches Committee for Division I men's lacrosse for the past two years as an advisor. He founded Laxpower, a web site that rated lacrosse teams from 1997 to 2019.}

\footnotetext[2]{Matthew Bomparola has played tennis for two decades and has worked as a coach and professional hitting partner. He received a degree from Princeton University in '21 and works as a data analyst, writer, and journalist.}

\footnotetext[3]{Lars Tiffany is the current head lacrosse coach at the University of Virginia, leading the team to national championships in ’19 and ‘21. He attended Brown University where he was team captain and later served as head coach.}

\section{Introduction}
Four years ago, the lacrosse Selection Criteria and Ranking Committee (abbreviated SCR, herein referred to as the "coaches' committee" or, simply, the "committee") set out to develop a rating system uniquely suited to select teams for the NCAA Men's Division I lacrosse championship tournament. The committee examined over a dozen potential methods, testing each algorithm against 20 years of data to assess its hypothetical at‑large selections.

Drawing on years of committee‑member experience, the history of sports ranking, and some of the existing academic literature, the committee identified a variety of desirable factors, including: \\

\newcounter{myroman}
\begin{list}{\Roman{myroman}.}{%
    \usecounter{myroman}%
    \setlength{\leftmargin}{1.5em}
    \setlength{\labelwidth}{1.5em}%
    \setlength{\labelsep}{0.5em}
    \setlength{\itemsep}{2pt}
    \setlength{\topsep}{0pt}
    \setlength{\parsep}{0pt}%
    \setlength{\partopsep}{0pt}%
}
  \item Emphasis on head‑to‑head and common opponents’ records instead of complex analytics.
  \item An unbiased, accurate adjustment for teams’ strength of schedule.
  \item The use of margins of victory.
  \item Minimizing bias and human intervention where possible.
  \item Common‑sense results.\\
\end{list}

The committee dreamed of a method that would replace the existing system for rating teams, which is highly based on the biased and mathematically flawed Ratings Percentage Index (RPI) and closed-door committees that - despite producing reasonable results - feel like a "byzantine black box," leading to a "large amount of speculation, second guessing, and debate each year about [committee] decisions." (\cite{Coleman2010}, \cite{Colley2002})

The committee finally settled on a method that
integrates a Sagarin‑inspired Power Rating\footnote{ Consult Appendix A for historical context on the development of Powerwise.} (which
measures team strength based on goal margins of
victory) and a variant of the pairwise comparison
method recently used by Division I Hockey. This
hybrid method aims to be simpler, more objective, and more transparent than the existing selection process, painting a clearer and fairer picture of which teams deserve to participate in the tournament given their performance in the regular season. This new method is known as "the Powerwise method" or "PWR," reflecting its pairwise and Power Rating components.

This document will clearly and simply describe the
Powerwise method and, with luck, convince you of
some of its merits, chief among them simplicity,
transparency, and common-sense results.

\begin{table}[H]
\centering
\caption{Comparison of Popular Methods}
\small
\resizebox{\columnwidth}{!}{%
\begin{tabular}{lcccc}
\toprule
Feature      & PWR   & RPI   & Committee & NPI \\ 
\midrule
W/L or Goal diffs.& Both & W/L & Both?    & W/L  \\ 
SoS Adjustment     & Yes & Yes**   & Yes? & Yes \\ 
Simple   & Yes    & Yes   & No    & No? \\ 
Transparent   & Yes    & Yes   & No    & Yes  \\ 
Objective    & Yes  & Yes    & No?    & Yes \\ 
Consistent & Yes & No & No? & No\\
Predictive & No* & No & No & No\\
Weak team penalty & No & Yes & No? & No\\
\bottomrule
\end{tabular}
}
\end{table}
\noindent\textit{*PWR's component Power Ratings are predictive.\\***The RPI SoS adjustment is flawed, see Appendix B.}

\section{The Powerwise Method}

\subsection{Simulating matchups with pairwise}
Powerwise works by simulating a scenario in which all teams in a division "play" against each other in hypothetical pairwise matches, competing for Powerwise points. 

For each matchup, one team "wins" and the other "loses" due to a three-step hierarchical procedure that evaluates, in order of importance: head-to-head results, each team's record against common opponents, and Power Ratings (to be discussed in the next section). 

Given a comparison between two hypothetical teams, the Powerwise point is awarded to:

\begin{enumerate}[label=\Roman*., leftmargin=1.5em, itemsep=2pt, parsep=0pt]
\item The team with the superior \textbf{head-to-head record.}\\
\textit{*Continue to step II only in case of a nonexistent or tied head-to-head record.}

\item The team with the superior record against \textbf{common opponents}.\footnote{Powerwise employs percentage win-loss when calculating common opponent records. One downside of this approach is that it risks downplaying repeated losses. This limitation is discussed in greater detail in Section 4.}\\
\textit{*Continue to step III only in case of a nonexistent, tied, or singular\footnote{The suggested Powerwise method optionally ignores any singular common opponent record, on the basis that the results of one game are likely tantamount to randomness. This "ad hoc" adjustment trades simplicity for a little more resistance to chance.} record against common opponents.}

\item The team with the higher \textbf{Power Rating.}
\end{enumerate}

After all matchups are examined, the teams are sorted from most-to-least deserving of an at-large bid on the basis of Powerwise points. For MD1 Lacrosse, 77 teams play (are compared in) 76 matchups, so the best possible Powerwise record for a given team is 76-0, and the worst is 0-76.\footnote{Appendix E presents an example Powerwise comparison. Appendix C presents an example ranking list.} 

A notable strength of the Powerwise method is that the majority of its pairwise comparisons are determined solely by on-field results in head-to-head and common opponents' games. Analysis of historical lacrosse data (2013-2024) finds just under 60\% of Powerwise points would have been awarded based on a superior on-field record. This enviable percentage falls as the number of teams in a division rises (assuming a constant number of games per season). 

\begin{table}[H]
\centering
\caption{Percentage of Powerwise comparisons determined by H2H and CO results}
\label{tab:powerwise-decisions}
\small
\resizebox{\columnwidth}{!}{%
  \begin{tabular}{@{}lccccc|c@{}}
    \toprule
    Season& 2013&2014&\(\dots\)& 2023 & 2024&Avg.*\\ 
    \midrule
    \% H2H decisive&21.5&19.6 &           &17.6&17.8& $\approx$19\% \\ 
    \% CO decisive    & 62.5 & 56.4 &           & 45.2 & 45.1 & $\approx$53\% \\ 
    \bottomrule
  \end{tabular}%
}
\end{table}
\noindent\textit{*Avg. Decisiveness of on-field (H2H or CO) = 58.6\%}
 
\subsection{Calculating Power Ratings}
As mentioned, in Powerwise matchups where on-field results are indecisive, Power Ratings break the tie.

Powerwise's Power Rating (PR) system is based on goal differentials and designed to estimate the average margin of victory between two teams on a neutral field. The formula is a bit mathematical for a general audience, but it's simpler than it looks.\footnote{$PR_i - PR_j$ = difference in estimated Power Ratings; $score_i - score_j$ = real difference in scores; $hfa$ = home-field advantage}

\begin{equation}\label{eq:power-rating}
  \sum_{i=0}^n \sum_{j=0}^m (PR_i - PR_j)
    = \sum_{i=0}^n \sum_{j=0}^m (score_i - score_j)
    \pm hfa
\end{equation}

Essentially, calculating PRs involves solving a large system of nonlinear equations based strictly on game scores and an adjustment for the home-field advantage, iterating until the average difference between two teams' Power Ratings is equal to the expected (or real) difference in scores were those two teams to play, adjusting for the home-field advantage.\footnote{Two ways to account for the home-field advantage include a constant adjustment across all teams in a division (determined by historical data or by some other means) or a team-by-team measure that accounts for the fact that different fields (say, Denver's mile-high stadium) perform differently. We opt for the simpler division-by-division average to simplify this analysis, but the preferred method is up for debate.}

Power Ratings implicitly account for strength of schedule thanks to the inclusion of goal differentials. To enable the use of goal differentials, PRs employ an adjustment to measured score data intended to disincentivize "running up the score." This merits further discussion.

\subsubsection{On running up the score}

Landmark studies conducted by Barrow et al. and Annis and Craig show that score differential-based methods tend to be both more predictive and more likely to converge on an interpretable solution than those that use only win-loss data. (\cite{Annis2005}, \cite{Barrow2013a})

While it might be unsurprising that the information contained in scores improves ratings, many methods still exclude score data, including Wesley Colley’s Colley Matrix. Colley justifies his use of win-loss ratios as a means to “keep it simple,” explaining that he didn’t want his method to require “ad hoc adjustments” to adjust for “runaway scores.” \cite{Colley2002}

Placing emphasis on simplicity is commendable but simplicity shouldn’t be bought at the expense of accuracy. NCAA lacrosse teams play far too few games per season—analysts just can’t afford to throw out valuable information. Powerwise is designed to use as few “ad hoc” adjustments as possible to enable the use of information-dense goal differentials while minimizing the risk of poor sportsmanship—i.e., incentives to run up huge scores against lesser opponents.\footnote{In 2001, famously, the Bowl Championship Series instructed its computer rankers to do away with statistics that measure goal differences altogether or to implement an ad hoc adjustment limiting score margins to plus-or-minus 21 in their code. Powerwise employs a similar adjustment.}

The ad hoc adjustment employed by this version of Powerwise places a “cap” on measured margins of victory at +7 or -7 goals. Any real-world goal that exceeds this limit is disregarded for the purposes of rankings. The number 7 was picked intuitively and may be adjusted by the committee as needed or for other sports.\footnote{An alternative method potentially applicable to future iterations of Powerwise involves normalizing goal differential data by taking the “nth” root with n being, for instance, some well-known normalizing constant. Once normalized, additional goals would be subject to diminishing returns.}

\subsubsection{"Why not the RPI?"}
The closest analogue to Powerwise is the pairwise ranking system used by the NCAA Men’s Division I Hockey committee, which uses an ancient method called the RPI in place of Powerwise’s Power Ratings. The RPI is famously problematic, not least because its win-loss methodology produces a systematic bias against mid-major and small conferences. Other problems with the RPI are detailed in Appendix B, including proof of a hypersensitivity to inconsequential results.

\section{Powerwise vs. Historical At-Large Picks}

\subsection{Data}
Official at-large and automatic qualifier data are pulled from the official NCAA websites (NCAA selection committee) and other readily-available sources. 

Game-level data come from laxpower.com (historical) and laxnumbers.com (more recent) and are processed to generate ranking lists for each method and the following analysis.

\subsection{Discrepancies}
Differences between historical selections and hypothetical Powerwise picks\footnote{Appendix C presents 5 years (skipping COVID-19 years) of historical and Powerwise MD1 lacrosse at-large selections.} are presented in Table 3 and include: 
\begin{itemize}[topsep=0pt, partopsep=0pt, parsep=0pt, itemsep=2pt]
    \item  In 2022, Powerwise prefers Notre Dame and Duke to committee picks of Cornell and Brown. 
    \item In 2019, Powerwise prefers Cornell and Denver to Notre Dame and Johns Hopkins.
    \item In 2018, Powerwise prefers Rutgers, Penn State and Bucknell to Syracuse, UVA, and Villanova. 
\end{itemize}

\begin{table}[H]
  \centering
  \small
  \caption{Powerwise vs.\ Official At‑Large Picks\\ (2013-2024, excluding years w/o discrepancies)}
  \label{tab:picks-2014-22}
  \begin{tabular*}{\columnwidth}{@{\extracolsep{\fill}} l l l @{}}
    \toprule
    Season & Powerwise picks       & Official picks        \\ 
    \midrule
    2022   & NOD, DUK              & COR, BRO              \\ 
    2019   & COR, DEN              & NOD, JHU              \\ 
    2018   & RUT, PSU, BUK         & SYR, UVA, VIL         \\ 
    2017   & DUK                   & UNC                   \\ 
    2016   & VIL, STB              & JHU, NAV              \\ 
    2015   & COR, HOF              & OSU, BRO              \\ 
    2014   & YAL                   & HAR                   \\ 
    \bottomrule
  \end{tabular*}
\end{table}

In ’23 and ‘24 Powerwise agrees with the committee other than in shuffling some seeding. The small number of discrepancies between Powerwise (a purely statistical method) and the committee’s expert- and RPI-informed approach might be surprising considering that "bubble" teams (those on the cusp of receiving an at-large bid) are often extraordinarily well-matched in a crowded field. 

That being said, while there is no "holy grail"\footnote{Colley’s extremely well-written analysis points out the “treachery” of checking an objective method against subjective truths. Answering whether a method performs well is remarkably difficult and out of scope for this whitepaper. \cite{Colley2002}} against which to compare Powerwise and historical at-large picks, there are good reasons to believe that Powerwise produces common sense ranking lists and potentially even selects teams that show stronger performance against tougher regular-season opponents than those selected by the current method. 

At the least, our analysis shows that Powerwise is much better suited to aid the tournament selection committee than the RPI. 

\subsubsection{Quantitative improvements}
In-depth statistical analysis is outside the scope of this introductory look at Powerwise, but an analysis inspired by Stocks-Smith's 2021 overview of their College Basketball Rating (CBR) is presented in Appendix F. \cite{StocksSmith2021}

Figure F2 plots game scores (y-axis) against opponents' strength as measured by Powerwise (on the x-axis), revealing that Powerwise's picks tended to outperform official picks in the preceding regular season. These results are statistically significant, stand up to the implementation of a goal cap of $\pm7$, and are robust to measuring opponent strength with RPI instead of Powerwise. 

\subsubsection{Qualitative improvements}
Powerwise meets the ideal characteristics set out by the coaches’ committee, including mathematical integrity, accuracy, objectivity, simplicity, transparency, consistency, versatility, and sportsmanship (with the caveat that Powerwise’s Power Ratings require an ad hoc or asymptotic adjustment to avoid incentivizing running up the score). 

The committee method, while admirably accurate given the difficulty of divining (or defining) deservedness of an at-large bid, falls short on simplicity, consistency, and transparency and, when informed by the RPI, appears to be less accurate, less objective, and more prone to biases with respect to Powerwise.\footnote{Appendix B further details problems we've identified with the RPI. One problem mentioned therein—systematic bias against smaller conferences—is clearly described and analyzed by Paul and Wilson in their 2015 publication. \cite{PaulWilson2015}} 

\section{Potential Limitations}
Like any method, Powerwise comes with potential complications. The following merit further discussion.\\

\begin{enumerate}[label=\Roman*., topsep=0pt, partopsep=0pt, parsep=0pt, itemsep=2pt]
    \item  Power Ratings are adjusted to account for the home-field advantage, but head-to-head and common opponent records are not. 
    \item Victory margins—i.e., goal differentials—may connote point spreads or sports gambling.
    \item Powerwise often leads to ties between teams with similar results. 
    \item Percentage win-loss records for common opponent matchups downplay repeated losses. \\
\end{enumerate}

Regarding each: \\

I.	Power Ratings are adjusted to account for the home-field advantage, but head-to-head and common opponent records are not. \\

 Should, say, a one-goal spread in a single common opponent game played away determine a Powerwise point? After all, teams tend to overperform at home and underperform in away games. Powerwise elects to ignore the home-field advantage in pairwise comparisons for the sake of simplicity and comprehensibility. \\

II.	Victory margins—i.e., goal differentials—may connote point spreads or sports gambling.\\
    
Methods that use game scores (i.e., goal differences or margins of victory) share some terminology with sports gambling—especially the phrase "point spread." Some NCAA sports try to avoid such connotations, especially given the increasing problem sports gambling poses to college campuses.\footnote{This is up to the discretion of the relevant NCAA committees, but it’s not clear whether avoiding the language associated with point spreads would decrease the incidence of sports gambling—the cat seems well and truly out of the bag.}

Powerwise puts connotations aside, noting that Vegas/gambling oddsmakers are highly incentivized to pick solid rankings and, for good reason, use margin of victory data in their calculations. The NCAA cannot afford to overlook a method that promises to generate legitimate results, especially considering arguments like those made by Nick Saban in 2023.\footnote{“When they told me we would be favored against three out of the four teams that got in the playoff, I’m like, why aren’t we in the playoffs?”—Nick Saban, in reference to the Las Vegas oddsmakers’ picks. \cite{Washington2023}}\\

III.	Powerwise often leads to ties between teams with similar results. \\

Appendix C displays a full list of Powerwise rankings for the MD1 2024 season. The ratings feature multiple ties between teams that equal numbers of Powerwise points during the pairwise analysis. Though Powerwise might seem less conclusive than other methods that produce fewer ties, the existence of ties amounts to recognizing that many lacrosse teams are so evenly matched that analytics alone are not able to accurately determine which deserves the higher rank. Appendix D resolves this problem by outlining procedures to differentiate between two-, three-, and four- (or more) team ties.\footnote{Alternative tie-breakers may be suggested as part of the Powerwise adoption process for different sports.}\\

IV.	Percentage win-loss records for common opponent matchups downplay repeated losses.\\

Consider a scenario involving Yale, Princeton, and Canisus. Say Yale loses to Princeton twice in close games and also say that Canisius, a team generally considered weaker than Yale, loses to Princeton only once. Should Canisius win its pairwise comparison against Yale on the basis that Yale has a worse record against their one common opponent? 

As-is, Powerwise considers Yale and Canisius’ on-field results as equal—0\% against their only common opponent. Lacking head-to-head results, the Powerwise point would likely be awarded to Yale based on its bigger Power Rating. An alternative, numeric, win-loss comparison test—i.e., Canisius’ record of 0-1 is better than Yale’s of 0-2 against Princeton—would reverse this result, potentially altering Yale’s end-of-season tournament chances.

Awarding the point to Canisius seems to fly in the face of logic—it’s obvious that Yale is the superior team. On the other hand, shouldn’t Yale be punished for their additional loss to Princeton? For simplicity’s sake and in favor of common sense, Powerwise prefers percentages over numeric records, deferring to Power Ratings slightly more often. 

\section{Conclusions or "Why Adopt Powerwise?"}
Powerwise has been rigorously developed, tested, and reviewed to fit lacrosse’s unique competitive environment. Substituting Powerwise for the current system of picking at-large candidates for the NCAA Men’s Division I lacrosse championship tournament would produce the following results: \\

I.	Increased trust from fans, coaches, and players in the post-season selection process as a result of increased simplicity and transparency. 

II.	Increased acceptance (to the extent possible) of tournament selection decisions thanks to Powerwise’s common-sense approach.

III. Satisfaction from the greater lacrosse community that Powerwise’s only metric of success is winning games (and doing so convincingly). Under Powerwise, each team controls its destiny on the field.

IV.	Peace of mind for the selection committee once the flawed, win-loss-based RPI is swapped for the simple, goal differential-based Power Rating statistic. PRs will improve the quality and consistency of at-large picks and the likelihood of rating convergence. 

V.	Gratefulness on the part of teams from smaller/weaker conferences that would enjoy a (higher) fair chance of earning an at-large berth because Powerwise avoids the biased RPI strength of schedule calculation.\\ 

In conclusion, while most methods focus on predictive accuracy, overstating their ability to rank and forecast in an environment featuring sparse data and human biases, Powerwise places more emphasis on other criteria, including simplicity, fairness, transparency, and ensuring that each fan, player, and coach feels that their favorite team controls its destiny on the field.

\onecolumn
\nocite{*}
\bibliographystyle{plainnat}
\bibliography{references}

\begin{thebibliography}{13}
\providecommand{\natexlab}[1]{#1}
\providecommand{\url}[1]{\texttt{#1}}
\expandafter\ifx\csname urlstyle\endcsname\relax
  \providecommand{\doi}[1]{doi: #1}\else
  \providecommand{\doi}{doi: \begingroup \urlstyle{rm}\Url}\fi

\bibitem[Annis and Craig(2005)]{Annis2005}
D.~H. Annis and B.~A. Craig.
\newblock Hybrid paired comparison analysis, with applications to the ranking of college football teams.
\newblock \emph{Journal of Quantitative Analysis in Sports}, 1\penalty0 (1), 2005.
\newblock \doi{10.2202/1559-0410.1000}.
\newblock URL \url{https://doi.org/10.2202/1559-0410.1000}.

\bibitem[Barrow et~al.(2013)Barrow, Drayer, Elliott, Gaut, and Osting]{Barrow2013a}
D.~Barrow, I.~Drayer, P.~Elliott, G.~Gaut, and B.~Osting.
\newblock Ranking rankings: An empirical comparison of the predictive power of sports ranking methods.
\newblock \emph{Journal of Quantitative Analysis in Sports}, 9\penalty0 (2):\penalty0 187--202, 2013.
\newblock \doi{10.1515/jqas-2013-0013}.
\newblock URL \url{https://doi.org/10.1515/jqas-2013-0013}.

\bibitem[Chartier et~al.(2011)Chartier, Kreutzer, Langville, and Pedings]{Chartier2011}
T.~P. Chartier, E.~Kreutzer, A.~N. Langville, and K.~E. Pedings.
\newblock Sensitivity and stability of ranking vectors.
\newblock \emph{SIAM Journal on Scientific Computing}, 33\penalty0 (3):\penalty0 1077--1102, 2011.
\newblock \doi{10.1137/090772745}.
\newblock URL \url{https://doi.org/10.1137/090772745}.

\bibitem[Coleman et~al.(2010)Coleman, DuMond, and Lynch]{Coleman2010}
B.~J. Coleman, J.~M. DuMond, and A.~K. Lynch.
\newblock Evidence of bias in ncaa tournament selection and seeding.
\newblock \emph{Managerial and Decision Economics}, 31\penalty0 (7):\penalty0 431--452, 2010.

\bibitem[Colley(2002)]{Colley2002}
W.~N. Colley.
\newblock Colley’s bias free college football ranking method.
\newblock Technical report, Princeton University, 2002.
\newblock URL \url{https://arxiv.org/abs/cs/0208005}.

\bibitem[Massey(1997)]{Massey1997}
K.~Massey.
\newblock \emph{Statistical Models Applied to the Rating of Sports Teams}.
\newblock Bluefield College, 1997.

\bibitem[Ochieng et~al.(2022)Ochieng, London, and Kr{\'e}sz]{Ochieng2022}
P.~J. Ochieng, A.~London, and M.~Kr{\'e}sz.
\newblock A forward-looking approach to compare ranking methods for sports.
\newblock \emph{Information}, 13\penalty0 (5):\penalty0 Article 5, 2022.
\newblock \doi{10.3390/info13050232}.
\newblock URL \url{https://doi.org/10.3390/info13050232}.

\bibitem[Paul and Wilson(2015)]{PaulWilson2015}
R.~J. Paul and M.~Wilson.
\newblock Political correctness, selection bias, and the ncaa basketball tournament.
\newblock \emph{Journal of Sports Economics}, 16\penalty0 (2):\penalty0 201--213, 2015.
\newblock \doi{10.1177/1527002512465413}.
\newblock URL \url{https://doi.org/10.1177/1527002512465413}.

\bibitem[Shah and Wainwright(2016)]{ShahWainwright2016}
N.~B. Shah and M.~J. Wainwright.
\newblock Simple, robust and optimal ranking from pairwise comparisons.
\newblock arXiv preprint arXiv:1512.08949, 2016.
\newblock URL \url{https://doi.org/10.48550/arXiv.1512.08949}.

\bibitem[Stefani(2011)]{Stefani2011}
R.~Stefani.
\newblock The methodology of officially recognized international sports rating systems.
\newblock \emph{Journal of Quantitative Analysis in Sports}, 7\penalty0 (4), 2011.
\newblock \doi{10.2202/1559-0410.1347}.
\newblock URL \url{https://doi.org/10.2202/1559-0410.1347}.

\bibitem[Stocks-Smith(2021)]{StocksSmith2021}
J.~Stocks-Smith.
\newblock College basketball rating (cbr): A new body-of-work metric for ncaa tournament selection.
\newblock \emph{Journal of Sports Analytics}, 7\penalty0 (1):\penalty0 47--55, 2021.
\newblock \doi{10.3233/JSA-200457}.
\newblock URL \url{https://doi.org/10.3233/JSA-200457}.

\bibitem[Washington(2023)]{Washington2023}
Matthew Washington.
\newblock Saban: Creating parity in college football is much harder than nfl.
\newblock \url{https://thescore.com/ncaaf/news/2666552}, June 2023.
\newblock [Online; accessed 20 July 2025].

\bibitem[website(2018)]{LaxPower2018}
{LaxPower.com} website.
\newblock Laxpower.com website shuts down.
\newblock New York Sportswriters blog, August~30 2018.
\newblock URL \url{http://newyorksportswriters.org/blog/2018-08-30-laxpower-lacrosse-website-shuts-down.shtml}.

\end{thebibliography}

\appendix

\newpage
\section*{Appendix A: Historical Background}
Two individuals served as inspiration in the development of Powerwise. They are Arpad Elo, a Hungarian-American physics professor, and Jeff Sagarin, an MIT statistician. 

The Elo system was developed by Arpad Elo in 1939 for the purpose of rating chess players. It functions a bit differently to Powerwise, but at its core, Elo involves inferring performance from wins, losses, and draws against players of varying strengths, i.e., of varying Elo ratings. The difference in Elo scores between two players is interpretable as an estimate (probabilistic) of the multi-game score should they play. In developing Powerwise, its creators were inspired by both the interpretability of Elo ratings and the strength of the idea that underlies it: pairwise comparisons.

Jeff Sagarin, a graduate MIT statistician, is recognized for bringing Power Ratings to the forefront of many sports, including football and basketball. His results appeared in USA Today from the 1980s until recently and were employed by the NCAA Men’s Division I basketball championship tournament committee. His method was one of eight used for the college football Bowl Championship Series (BCS) from 1998 to 2014. When the committee eliminated point margins, Sagarin developed a second method without point spreads which remained in use by the BCS. 

His algorithm has never been publicly revealed, but it is known that Sagarin used a constant home-field advantage for football and basketball—for instance, 3 or more points, depending on the sport—and introduced cut-offs on margins of victory to discourage running up the score.

Since Sagarin Power Ratings produce “point spreads” and ranking lists that tend to be transitive, many gamblers use them to try to “beat the odds” set by oddsmakers. It is probably for this reason that Sagarin’s ratings have not been embraced by the NCAA, at least publicly.  

The Power Rating used by Powerwise was first developed by Dr. Lawrence Feldman for college lacrosse in 1997 (under the name of Laxpower) and wasn’t specifically guided by inside knowledge of Sagarin’s algorithm. The similarity in hypothetical results suggests that the algorithms are similar in function. Both use a constant home-field advantage and goal margin limit, for instance. Laxpower has been described as “uncannily accurate.” \cite{LaxPower2018}

It isn’t clear which convergence criteria is used in Sagarin’s rating algorithm, but Feldman iteratively solves equations for each team simultaneously until the average error for each rating is zero when summing the difference between Power Ratings and real margins of victory. 

While pairwise comparisons are unusually powerful, given that college sports ratings are unlike chess ratings in that they are limited to a dataset of one (small) season, additional criteria/metrics are required to achieve convergence. That’s where Power Ratings come in handy. Lawrence Feldman stuck Power Ratings into an innovative pairwise method and Powerwise was born.

\newpage
\section*{Appendix B: Three Problems With the Ratings Percentage Index}
The Rating Percentage Index (RPI) system for ranking teams in Division I sports, particularly lacrosse, has several flaws that affect its accuracy and reliability. Here are three major deficiencies illustrated by example cases:

\subsection*{Issue 1: Strength of Schedule (SOS) is Based on Opponent Performance, Not Team Performance}
The RPI’s SOS calculation reflects the win percentage of a team’s opponents rather than how a team performed against those opponents. Here’s a hypothetical scenario: Should Hampton (Men’s D1 lacrosse), ranked 76th in 2024, be invited to the ACC, scheduled tougher competition, and losing all games by a wide margin, despite its losses, Hampton's RPI would jump from 0.3179 to 0.5053, improving by 38 places due to the stronger schedule alone.

This suggests the RPI is more sensitive to a team's opponents’ performance rather than its own, distorting rankings. The formula's reliance on opponent win percentages (OWP) and opponents’ opponents win percentages (OOWP) creates a situation wherein a team can artificially rise in the rankings by merely playing tougher competition, regardless of actual performance. Table B1 presents the actual RPI results from the 2024 season. If instead, Hampton was to have played in the ACC with a different schedule, table B2 shows Hampton’s RPI improvement in rating and rank, from 0.3179 to 0.5053 or 76th to 38th, despite remaining the same team.

\begin{table}[H]
  \centering
  \small
  \caption{Hampton Dilemma: original vs.\ “ACC‑ified” RPI rankings}
  \begin{subtable}[t]{0.48\textwidth}
    \centering
    \caption{Original RPI Rankings}
    \label{tab:B1}
    \begin{tabular}{@{}r l c r r@{}}
      \toprule
      \textbf{RPI Rank} & \textbf{Team Name}   & \textbf{RPI} & \textbf{Wins} & \textbf{Losses} \\
      \midrule
      70 & Mt.\ St.\ Mary’s & 0.3663 &  1 & 14 \\
      71 & St.\ Bonaventure & 0.3538 &  1 & 11 \\
      72 & Wagner           & 0.3468 &  1 & 12 \\
      73 & Queens           & 0.3353 &  2 & 11 \\
      74 & Mass–Lowell      & 0.3350 &  0 & 12 \\
      75 & Lindenwood       & 0.3235 &  0 & 12 \\
      76 & \textbf{Hampton} & \textbf{0.3179} &  0 & 13 \\
      \bottomrule
    \end{tabular}
  \end{subtable}%
  \hfill
  \begin{subtable}[t]{0.48\textwidth}
    \centering
    \caption{Adjusted (ACC) Schedule RPI}
    \label{tab:B2}
    \begin{tabular}{@{}r l c r r@{}}
      \toprule
      \textbf{RPI Rank} & \textbf{Team Name}   & \textbf{RPI} & \textbf{Wins} & \textbf{Losses} \\
      \midrule
      38 & \textbf{Hampton} & \textbf{0.5053} &  0 & 13 \\
      39 & Vermont          & 0.4907          &  8 &  8 \\
      40 & Drexel           & 0.4883          &  5 &  9 \\
      41 & Air Force        & 0.4876          &  9 &  6 \\
      42 & Quinnipiac       & 0.4874          &  9 &  5 \\
      43 & Brown            & 0.4790          &  3 & 11 \\
      44 & LIU              & 0.4730          & 10 &  4 \\
      \bottomrule
    \end{tabular}
  \end{subtable}%
  \label{tab:hampton‑dilemma}
\end{table}

\noindent\textit{Note.} Hampton ranked 76th during the 2024 season. When its schedule is “ACC‑ified,” the same roster jumps to an RPI of 38.

\subsection*{Issue 2: Ignoring Goal Differentials Invites Inaccuracy}
The second major issue is that RPI only accounts for wins and losses, disregarding the margin of victory or defeat. Consider Team A (which defeats teams by large margins) versus Team B (which wins against the same teams by narrow margins). The RPI system would rate both teams similarly, despite Team A’s superior performance. This overlooks the fact that a decisive win (e.g., 15-0) indicates a stronger team than a close win (e.g., 13-12). The absence of goal differential data reduces the granularity of the analysis and leads to less accurate rankings. This approach is mathematically questionable, as it ignores key performance metrics that better reflect a team's strength.

\subsection*{Issue 3: Hypersensitivity to Irrelevant Games}
The RPI’s sensitivity to small changes in irrelevant or inconsequential games further compromises its accuracy. The example illustrated in Tables 5 and 6 shows how a remote upset, which has little bearing on tournament qualification, can significantly impact rankings. 

The left and right panels of Table 5 present ranking lists based on the same data, except for switching the result of a close game played early in the season between two relatively unimportant teams: Delaware and Lafayette. The right panel of Table 5 shows that the rankings of multiple top-20 teams have shifted despite Lafayette and Delaware having little direct interaction with the tournament contenders. Such hypersensitivity undermines the sense that teams control their own destiny. Table 6 repeats this experiment with Power Ratings, indicating less sensitivity to similar perturbations in the data. Only one, explainable, ranking change occurs in the right panel.

\begin{table}[H]
  \centering
  \small
  \caption{RPI rankings before vs.\ after a single‑game upset\\
           (teams whose rank changed are in \textbf{bold}}
  \label{tab:B3-full}
  \begin{tabular}{@{}r l c l c@{}}
    \toprule
    & \multicolumn{2}{c}{\textbf{Original RPI}} 
    & \multicolumn{2}{c}{\textbf{Perturbed RPI}} \\ 
    \cmidrule(lr){2-3}\cmidrule(lr){4-5}
    \textbf{Rank} & \textbf{Team}        & \textbf{RPI} 
                  & \textbf{Team}        & \textbf{RPI} \\
    \midrule
     1 & Notre Dame        & 0.7100 & Notre Dame        & 0.7097 \\
     2 & Duke              & 0.6632 & Duke              & 0.6631 \\
     3 & Johns Hopkins     & 0.6520 & Johns Hopkins     & 0.6517 \\
     4 & Syracuse          & 0.6404 & Syracuse          & 0.6381 \\
     5 & Virginia          & 0.6371 & Virginia          & 0.6368 \\
     6 & Denver            & 0.6246 & Denver            & 0.6246 \\
     7 & Maryland          & 0.6223 & Maryland          & 0.6223 \\
     8 & \textbf{Princeton}         & 0.6158 & \textbf{Penn State} & 0.6157 \\
     9 & \textbf{Penn State}        & 0.6158 & \textbf{Princeton}  & 0.6157 \\
    10 & Georgetown        & 0.6147 & Georgetown        & 0.6146 \\
    11 & Penn              & 0.6044 & \textbf{Cornell}    & 0.6038 \\
    12 & \textbf{Cornell}           & 0.6041 & \textbf{Yale}       & 0.6018 \\
    13 & \textbf{Michigan}          & 0.6019 & \textbf{Penn}       & 0.6018 \\
    14 & \textbf{Yale}              & 0.6017 & \textbf{Michigan}   & 0.5994 \\
    15 & St.~Joseph’s      & 0.5970 & St.~Joseph’s      & 0.5967 \\
    \bottomrule
  \end{tabular}
\end{table}

As you can see, flipping that one “irrelevant” result between Delaware and Lafayette scrambles teams' RPI-derived ranks—especially around the bubble and even in the top 10. The Power Rating list (follows) sees one adjacent swap.

\vspace{1em}

\begin{table}[H]
  \centering
  \small
  \caption{Power Ratings before vs.\ after the same upset\\
           (teams whose rank changed are in \textbf{bold}}
  \label{tab:B4-full}
  \begin{tabular}{@{}r l c l c@{}}
    \toprule
    & \multicolumn{2}{c}{\textbf{Original PR}} 
    & \multicolumn{2}{c}{\textbf{Perturbed PR}} \\ 
    \cmidrule(lr){2-3}\cmidrule(lr){4-5}
     \textbf{Rank} 
    & \textbf{Team}        & \textbf{PR} 
    & \textbf{Team}        & \textbf{PR} \\
    \midrule
     1  & Notre Dame      & 99.90 & Notre Dame      & 99.90 \\
     2  & Duke            & 97.51 & Duke            & 97.52 \\
     3  & Virginia        & 97.38 & Virginia        & 97.37 \\
     4  & Syracuse        & 97.02 & Syracuse        & 97.00 \\
     5  & Penn State      & 96.90 & Penn State      & 96.91 \\
     6  & Johns Hopkins   & 96.77 & Johns Hopkins   & 96.77 \\
     7  & Princeton       & 96.46 & Princeton       & 96.47 \\
     8  & Georgetown      & 95.78 & Georgetown      & 95.78 \\
     9  & Maryland        & 95.67 & Maryland        & 95.68 \\
    10  & Cornell         & 95.67 & Cornell         & 95.68 \\
    11  & Yale            & 95.47 & Yale            & 95.49 \\
    12  & Denver          & 95.47 & Denver          & 95.47 \\
    13  & Michigan        & 95.25 & Michigan        & 95.21 \\
    14  & \textbf{Towson} & 94.80 & \textbf{Army}   & 94.82 \\
    15  & \textbf{Army}   & 94.68 & \textbf{Towson} & 94.69 \\
    \bottomrule
  \end{tabular}
\end{table}

\subsection*{Conclusion}
The aforementioned issues with RPI—its flawed SOS calculation, its disregard for goal differentials, and its hypersensitivity to irrelevant games—compromise its effectiveness as a ranking system. As a result, the RPI fails to provide a true reflection of a team's ability, performance, and undermines the sense that teams control their own destiny. This evidence highlights the need for accurate and sophisticated systems that can incorporate additional data, such as goal differentials, and reduce the impact of arbitrary results. In general, Power Ratings do a better job of representing a team's competitive standing than does the RPI.

\newpage

\section*{Appendix C: Additional Tables and Figures}

\begin{table*}[ht]
  \centering
  \small
  \caption{Powerwise ratings for the 2024 season. \textbf{Tied ratings} are in bold.}
  \label{tab:powerwise-2024-4vert}
  \begin{tabular}{@{} 
      l r @{\hspace{0.6em}} 
      l r @{\hspace{0.6em}} 
      l r @{\hspace{0.6em}} 
      l r @{}}
    \toprule
    Team            & Rating 
  & Team            & Rating 
  & Team            & Rating 
  & Team            & Rating \\
    \midrule
    Notre Dame      & 74.00 & Vermont         & 36.00 & Harvard         & 56.00 & NJIT            & 20.00 \\
    Duke            & 72.00 & Lafayette       & \textbf{35.00} & Villanova       & 55.00 & Binghamton      & 19.00 \\
    Virginia        & 71.00 & Quinnipiac      & \textbf{35.00} & Delaware        & 54.00 & Bellarmine      & 18.00 \\
    Syracuse        & 70.00 & Sacred Heart    & \textbf{35.00} & Richmond        & 53.00 & St Johns        & 18.00 \\
    Johns Hopkins   & \textbf{69.00} & Drexel          & 34.00 & Colgate         & \textbf{51.00} & Cleveland St    & 15.00 \\
    Penn State      & \textbf{69.00} & Air Force       & \textbf{32.00} & Loyola          & \textbf{51.00} & VMI             & 13.00 \\
    Maryland        & 68.00 & Stony Brook     & \textbf{32.00} & Rutgers         & \textbf{51.00} & Holy Cross      & 12.00 \\
    Denver          & \textbf{67.00} & UMBC            & 30.00 & Boston Univ     & \textbf{50.00} & Le Moyne        & 12.00 \\
    Georgetown      & \textbf{67.00} & Dartmouth       & \textbf{28.00} & Lehigh          & \textbf{50.00} & Canisius        & 11.00 \\
    Cornell         & \textbf{66.00} & Marquette       & \textbf{28.00} & Navy            & 48.00 & Robert Morris   &  9.00 \\
    Princeton       & \textbf{66.00} & Hofstra         & 27.00 & Utah            & 45.00 & Detroit Mercy   &  8.00 \\
    Yale            & 65.00 & Monmouth        & 26.00 & Providence      & 44.00 & St Bonaventure  &  8.00 \\
    Michigan        & 62.00 & Hobart          & 25.00 & Bryant          & \textbf{43.00} & Mercer          &  7.00 \\
    Army            & \textbf{61.00} & Bucknell        & 24.00 & High Point      & \textbf{43.00} & Mt St Marys     &  5.00 \\
    Towson          & \textbf{61.00} & LIU             & 23.00 & Jacksonville    & 41.00 & Queens          &  4.00 \\
    North Carolina  & 60.00 & Manhattan       & \textbf{21.00} & Albany          & 40.00 & Wagner          &  4.00 \\
    Penn            & 60.00 & Marist          & \textbf{21.00} & Brown           & 39.00 & UMass‑Lowell    &  2.00 \\
    St Joseph's     & 58.00 & Merrimack       & \textbf{21.00} & Massachusetts   & 38.00 & Lindenwood      &  1.00 \\
    Ohio State      & 57.00 & Siena           & \textbf{21.00} & Fairfield       & 36.00 & Hampton         &  0.00 \\
    \bottomrule
  \end{tabular}
  
  \vspace{0.5ex}
  \footnotesize
  \noindent\textit{Note:} Ties are ironed-out using the common-sense tie-breaker procedures described in Appendix D.
\end{table*}

\begin{table*}[ht]
  \centering
  \caption{At-Large Picks by Season: Official vs. Powerwise}
  \label{tab:atlarge-final}
  \small
  
  \newcolumntype{R}{>{\raggedleft\arraybackslash}p{0.5cm}}
  
  \begin{subtable}[t]{0.31\textwidth}
    \centering
    \caption*{\textbf{Season 2024}}
    \begin{tabular}{@{}Rll@{}}
      \toprule
      \# & Official & Powerwise \\
      \midrule
      1 & Notre Dame     & Notre Dame \\
      2 & Duke           & Duke \\
      3 & Johns Hopkins  & \textit{Virginia} \\
      4 & Syracuse       & Syracuse \\
      5 & Denver         & \textit{Johns Hopkins} \\
      6 & \textit{Virginia} & Penn State \\
      7 & \textit{Maryland} & Maryland \\
      8 & Penn State     & \textit{Denver} \\
      \bottomrule
    \end{tabular}
  \end{subtable}%
  \hfill
  \begin{subtable}[t]{0.31\textwidth}
    \centering
    \caption*{\textbf{Season 2023}}
    \begin{tabular}{@{}Rll@{}}
      \toprule
      \# & Official & Powerwise \\
      \midrule
      1 & Duke          & \textit{Notre Dame} \\
      2 & Virginia      & Virginia \\
      3 & \textit{Notre Dame} & Duke \\
      4 & Maryland      & \textit{Cornell} \\
      5 & Penn State    & Maryland \\
      6 & Johns Hopkins & Penn State \\
      7 & \textit{Cornell} & Johns Hopkins \\
      8 & Yale          & Yale \\
      \bottomrule
    \end{tabular}
  \end{subtable}%
  \hfill
  \begin{subtable}[t]{0.31\textwidth}
    \centering
    \caption*{\textbf{Season 2022}}
    \begin{tabular}{@{}Rll@{}}
      \toprule
      \# & Official & Powerwise \\
      \midrule
      1 & Yale        & Virginia \\
      2 & Princeton   & \textit{Notre Dame}$^{\dagger}$ \\
      3 & Rutgers     & Princeton \\
      4 & Cornell$^{\dagger}$ & Rutgers \\
      5 & Brown$^{\dagger}$ & \textit{Duke}$^{\dagger}$ \\
      6 & Virginia    & Yale \\
      7 & Harvard     & Harvard \\
      \bottomrule
    \end{tabular}
  \end{subtable}

  \vspace{2ex}

  \begin{subtable}[t]{0.31\textwidth}
    \centering
    \caption*{\textbf{Season 2019}}
    \begin{tabular}{@{}Rll@{}}
      \toprule
      \# & Official & Powerwise \\
      \midrule
      1 & Duke           & Yale \\
      2 & Virginia       & Duke \\
      3 & Yale           & \textit{Loyola} \\
      4 & \textit{Notre Dame}$^{\dagger}$ & Virginia \\
      5 & Loyola         & Syracuse \\
      6 & Syracuse       & \textit{Cornell}$^{\dagger}$ \\
      7 & Johns Hopkins$^{\dagger}$ & \textit{Maryland} \\
      8 & Maryland       & \textit{Denver}$^{\dagger}$ \\
      \bottomrule
    \end{tabular}
  \end{subtable}%
  \hfill
  \begin{subtable}[t]{0.31\textwidth}
    \centering
    \caption*{\textbf{Season 2018}}
    \begin{tabular}{@{}Rll@{}}
      \toprule
      \# & Official & Powerwise \\
      \midrule
      1 & Maryland    & Yale \\
      2 & Yale        & Duke \\
      3 & Duke        & Maryland \\
      4 & \textit{Notre Dame} & Denver \\
      5 & Syracuse$^{\dagger}$ & \textit{Rutgers}$^{\dagger}$ \\
      6 & Denver      & \textit{Penn State}$^{\dagger}$ \\
      7 & Virginia$^{\dagger}$ & \textit{Bucknell}$^{\dagger}$ \\
      8 & Villanova$^{\dagger}$ & \textit{Notre Dame} \\
      \bottomrule
    \end{tabular}
  \end{subtable}%
  \hfill
  \begin{subtable}[t]{0.31\textwidth}
    \phantom{.}
  \end{subtable}

  \vspace{1ex}
  \footnotesize
  \noindent\textit{Note:} Italicized teams indicate position changes between Official and Powerwise selections. 
  $^{\dagger}$Indicates teams that appear in one list but not the other (i.e., selection disagreements).
\end{table*}

\clearpage 

\newpage
\section*{Appendix D: Powerwise Tie-breaker Procedures}
The Powerwise Tie-Breaker Procedure outlined below provides a detailed system for breaking ties between teams with identical Powerwise numbers. The procedure takes a step-by-step approach depending on how many teams are tied and uses: head-to-head results, records against common opponents, and power ratings. Here's a breakdown of the tie-breaker process:\\

\begin{center}
    \includegraphics{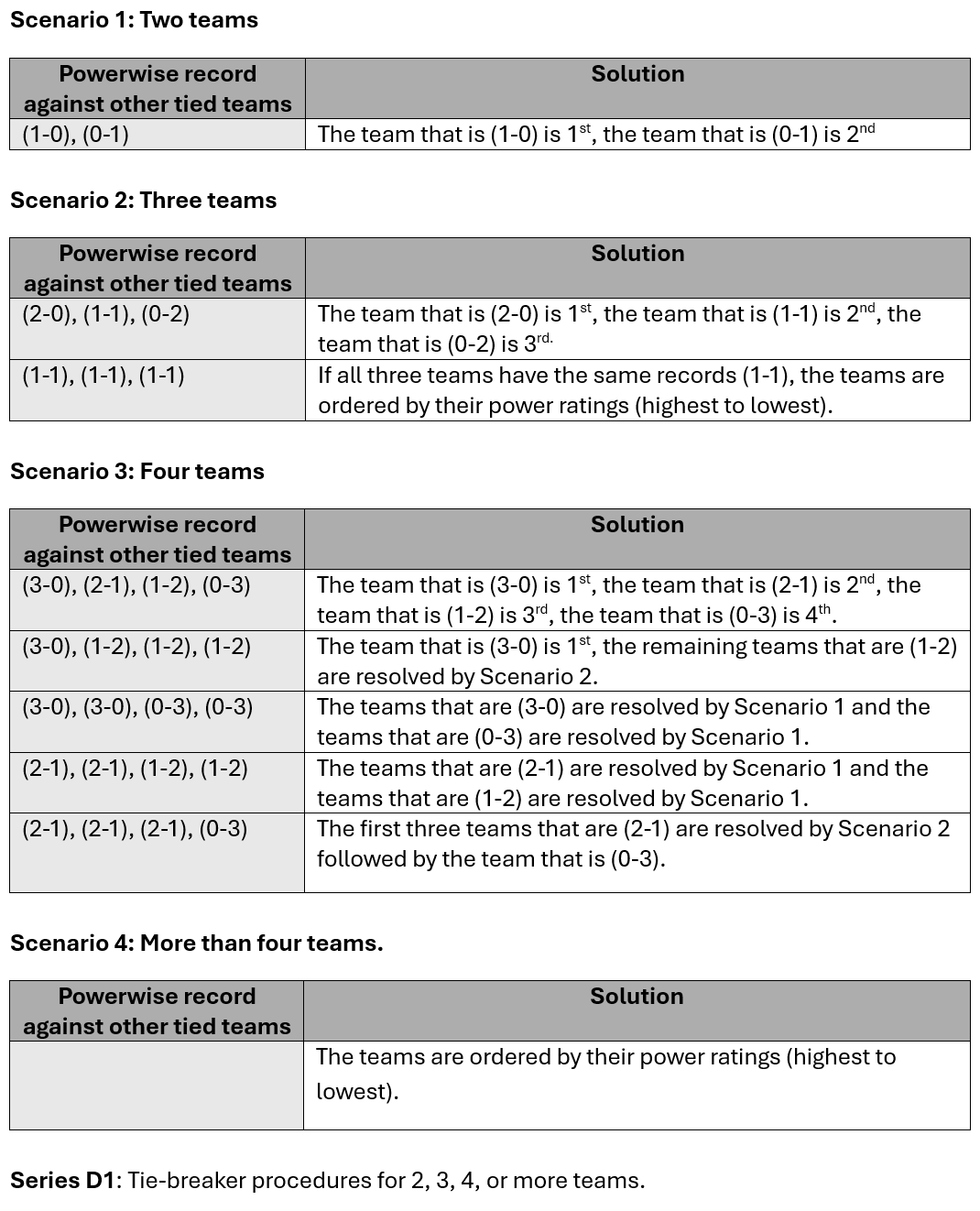}
\end{center}

\newpage
\section*{Appendix E: An Example Powerwise Scenario}
Consider, as an example, Powerwise matchups for Yale illustrated in Table E1.\\\\
I.	Yale is awarded the win over Brown because it went 1-0 against Brown in a head-to-head matchup.\\
II.	Yale did not play Delaware but had a better record than Delaware in games against common opponents (2-0) versus (1-1). If Delaware had gone (2-0) they both would have tied because the comparison is based on percentage wins.\\
III.	Yale and Richmond did not meet, nor did they have any common opponents, meaning that the Powerwise point is awarded to Yale due to its higher power rating.\\

\begin{center}
    \includegraphics{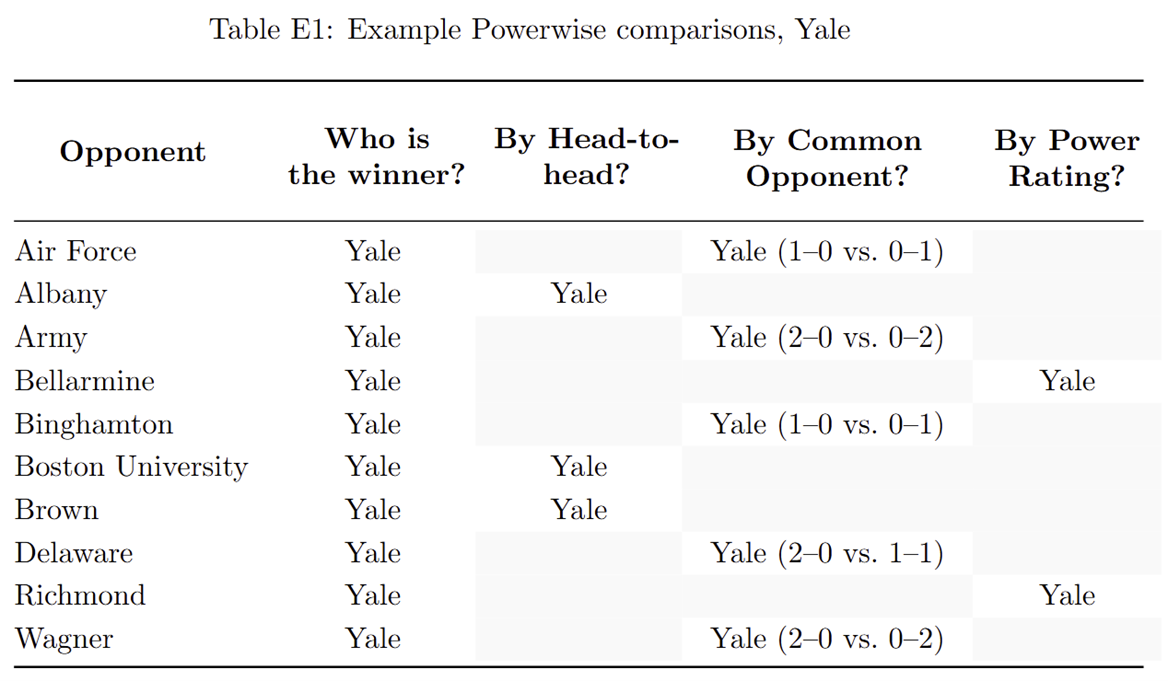}\\
    \textit{Note:} Extending this example to all of Yale’s comparisons (for this season), 14 were settled by head-to-heads, 37 by common opponents’ games, and 25 were determined by Power Ratings.
\end{center}

\newpage
\section*{Appendix F: Statistical Analysis}
The main finding presented in this statistical appendix is presented in Figure F2, namely that aggregating game-level data for all discrepancies in at-large selection between Powerwise's ranking list and the historical record, Powerwise appears to pick teams that are slightly more deserving given their regular season on-field performance—a result that's robust within a 95\% confidence interval and holds up to a plus-or-minus 7 goal cap.

Table F1 rejects the null hypothesis, presenting evidence that Powerwise’s ranking lists are especially unique around the at-large cutoff or “bubble,” and Table F3 report the discrepancies between Powerwise’s and historical official at-large tournament selections. Finally, Figures F4 and F5 illustrate that F2’s findings hold up across seasons (at least at the high end of the opponent’s strength spectrum) and when implementing a goal margin of victory cutoff of 7. \\

\begin{center}
\includegraphics{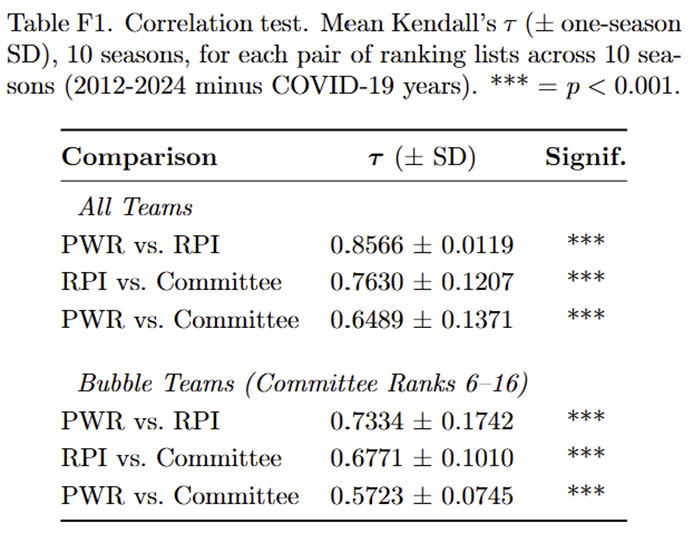}\\
\end{center}

\textbf{Table F1} presents Kendall's $\tau$ correlation coefficients comparing Powerwise (PWR), RPI, and Committee rankings across 10 seasons of data (2012-2024 minus COVID-19 years). While PWR and RPI show strong overall correlation ($\tau$ = 0.8566, p < 0.001), this relationship weakens considerably for bubble teams ranked 6-16 ($\tau$ = 0.7334, p < 0.001), indicating that Powerwise produces substantively different rankings where it matters most for tournament selection. 

\textbf{Figure F2} is based on an analysis performed by Stocks-Smith and demonstrates that Powerwise selections outperform official picks by a consistent margin across all opponent strengths (significant at p < 1.25 × $10\textsuperscript{-15}$).\cite{StocksSmith2021} The parallel regression lines indicate this advantage is uniform rather than opponent-dependent, suggesting Powerwise identifies teams with superior overall performance rather than favorable matchup characteristics. This 1-2 goal per game advantage persists across the full range of opponent quality.

\begin{center}
\includegraphics{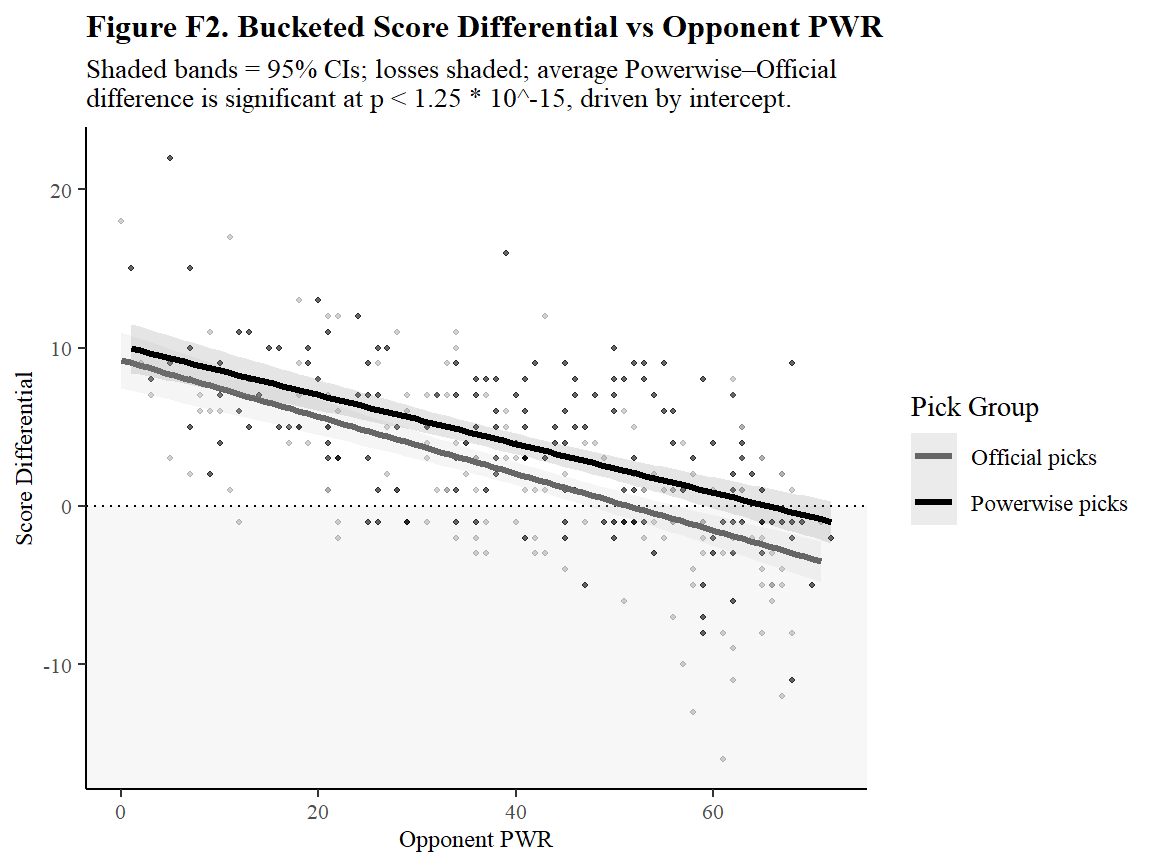}
\end{center}

\textbf{Table F3} reveals that across ten seasons, Powerwise and official selections disagreed on 13 at-large bids, with discrepancies ranging from complete agreement (2023-24) to divergence on three selections (2018). 

\textbf{Figure F4} displays the results of the same calculation shown in F2 for all discrepant years. The shaded confidence intervals indicate the high level of uncertainty associated with a single season’s worth of game results. Despite dubious statistical significance at these low sample sizes, Powerwise teams outperform the Official picks in many of the years analyzed, at least against teams at the higher end of the opponent strength spectrum. 

Finally, \textbf{Figure F5} demonstrates that the results displayed in F2 hold up to measuring opponent strength with the RPI (instead of PWR) and to an adjustment to combat “running up the score,” that is, capping the measured goal difference of any given game at plus or minus seven goals. 

\begin{table}[ht]
\centering
\captionsetup{labelformat=empty}
\caption{Table F3. Powerwise vs.\ Official At-Large Picks by Season}
\label{tab:atlarge-picks}
\begin{tabular}{l l l}
\toprule
Season & Powerwise picks                  & Official picks                     \\
\midrule
2022   & Notre Dame, Duke                 & Cornell, Brown                     \\
2019   & Cornell, Denver                  & Notre Dame, Johns Hopkins          \\
2018   & Rutgers, Penn State, Bucknell    & Syracuse, Virginia, Villanova      \\
2017   & Duke                             & North Carolina                     \\
2016   & Villanova, Stony Brook           & Johns Hopkins, Navy                \\
2015   & Cornell, Hofstra                 & Ohio State, Brown                  \\
2014   & Yale                             & Harvard                            \\
\bottomrule
\end{tabular}
\end{table}

\begin{center}
    \includegraphics{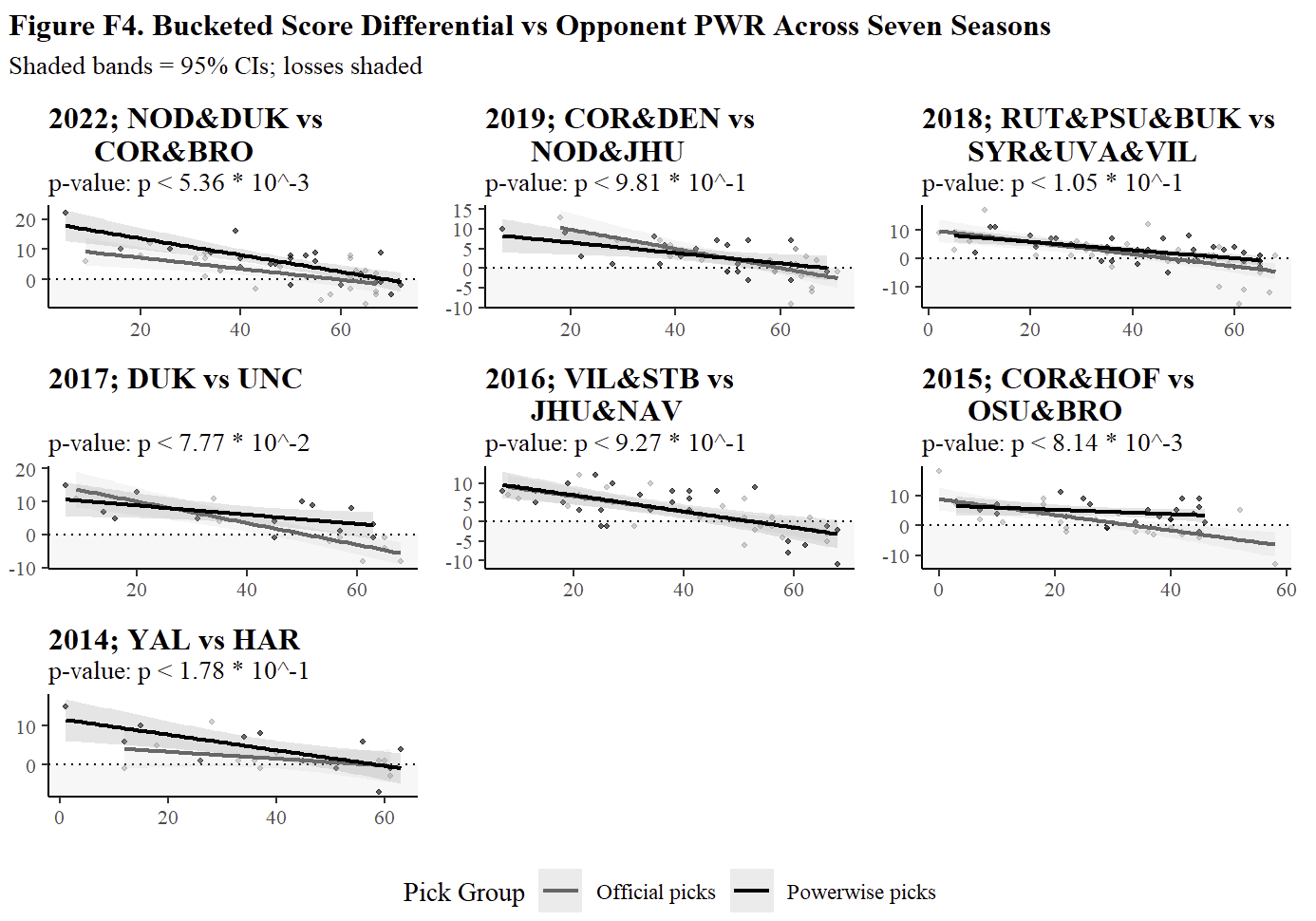}
    \includegraphics{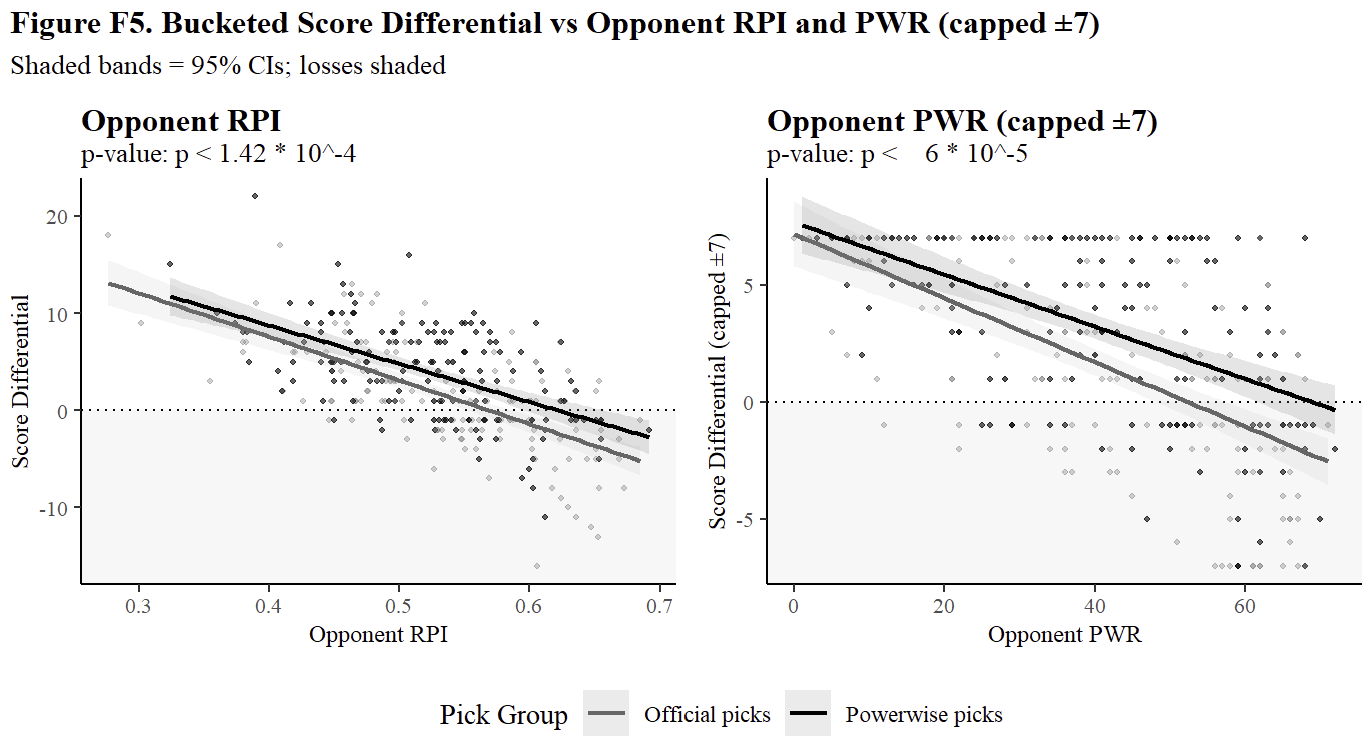}
\end{center}

\end{document}